\documentclass[preprint,proceedings]{rmaa}

\suppressfulladdresses 

\usepackage{paralist}

\usepackage{psfrag,color}

\newcommand{\kms}{\ensuremath{{\rm km\,s^{-1}}}}                   
\newcommand{\msunyr}{\ensuremath{\mathit{M}_{\odot}{\rm yr}^{-1}}}   
\newcommand{\rsun}{\ensuremath{\mathit{R}_{\odot}}}                  

\newcommand{\mdot}{\ensuremath{\dot{M}}}                             
\newcommand{\rstar}{\ensuremath{\mathit{R}_{\star}}}                 
\newcommand{\teff}{\ensuremath{\mathit{T}_{\rm eff}}}                
\newcommand{\vinf}{\ensuremath{v_{\infty}}}                          

\SetYear{2006}
\SetConfTitle{Massive Stars Argentina}

\title{LBV\lowercase{s} and the nature of the S Dor cycles: the case of AG~Carinae} 

\author{
  J. H. Groh,\altaffilmark{1} 
  A. Damineli,\altaffilmark{1}
  and D. J. Hillier\altaffilmark{2}}

\altaffiltext{1}{IAG - Universidade de S\~ao Paulo, Rua do Mat\~ao 1226, Cid.
Universit\'aria, CEP 05508-900, S\~ao Paulo SP, Brazil (groh, damineli@astro.iag.usp.br).}

\altaffiltext{2}{Physics and Astronomy Department, University of Pittsburgh,
3941 O'Hara Street, Pittsburgh PA 15260, USA (hillier@pitt.edu) .}

\shortauthor{Groh, Damineli, \& Hillier}
\shorttitle{AG Car and the S Dor cycles}

\listofauthors{J. H. Groh, A. Damineli, \& D. J. Hillier}
\indexauthor{Groh, J. H.}
\indexauthor{Damineli, A.}
\indexauthor{Hillier, D. J.}

\abstract{We present the results of a detailed spectroscopic analysis of 20 years of observations of 
AG Carinae using the radiative transfer code CMFGEN. Among the conclusions of this work, we highlight the
importance of including time-dependent effects in the analysis of the full S Dor cycle. We obtained that the
mass-loss rate is approximately constant during the cool-phases, implying that the S Dor-type eruptions begin
well earlier than the maximum seen in the visual lightcurve. We also determined that the S Dor cycles are
ultimately a consequence of an increase/decrease of the hydrostatic radius in combination with the formation
of a pseudo-photosphere.  }

\addkeyword{Stars: individual: AG Car}
\addkeyword{Stars: early-type}
\addkeyword{Stars: mass loss}

\begin{document}
\maketitle

\section{AG Car in context}
\label{intro}

The massive star AG Carinae has been the most variable object among
the Galactic luminous blue variables during the last 20 years. Dramatic
photometric changes of 3 magnitudes in the $V$ band have been recorded, while the stellar spectra
have indicated effective temperatures ranging from about 9kK to 25 kK.
AG Car is a rare member of the LBV class that shows strong S Dor-type variations
and is surrounded by a bipolar nebula.

This work is a result of a monitoring campaign spanning 20 years of observations
of AG Car to obtain its stellar and wind parameters for each epoch. The first results were discussed in Groh
et al. (2006), highlighting the first observational evidence of fast rotation in LBVs.
The evolution of both stellar (effective temperature, radius, luminosity) and
wind parameters (mass-loss rate and wind terminal velocity) along the S Dor
cycle, and the correlations among them, provides significant insights about the
physical processes going on during this short and unstable evolutionary stage.

\section{Spectroscopic modeling}
\label{model}

The physical parameters of AG Car were obtained through spectroscopic modeling 
using the radiative transfer code CMFGEN (Hillier \& Miller 1998), which allows a 
simultaneous solution of the radiative transfer and rate equations. Spherical
geometry and steady state are assumed to solve those equations, while clumping is included using a
depth-variable filling factor. Fully-blanketed models were calculated for AG Car,
taking into account the presence of over than 50,000 transitions of H, He, C, N,
O, Si, P, Na, Mg, Fe, Mn, Co and Ni.

\begin{figure}[!h]
  \includegraphics[width=\columnwidth]{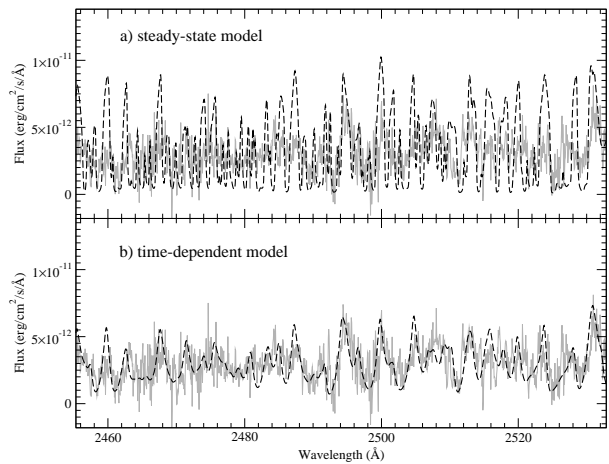}
  \caption{Comparison between the ultraviolet spectrum of AG Car
  taken in 1991 October (grey) and a steady-state model (panel $a$, dashed) and a
  time-dependent model (panel $b$, dashed). The better quality of the time-dependent fits is remarkable.}
  \label{uvtimedep}
\end{figure}

For the first time in the analysis of LBV winds, time-dependent effects were included consistently in the
velocity law and density structure of the wind assuming that the mass-loss rate and wind terminal velocity
change linearly as a function of time (J. H. Groh et al. 2007, in preparation). Significant differences can be seen in
the fits obtained with a time-dependent model in comparison with a steady-state model, especially in the ultraviolet
region (Fig.~\ref{uvtimedep}).

\section{Results and conclusions}
\subsection{Comparison between the minimum epochs}

The stellar and wind parameters are different during consecutive minima as derived from the analysis of
spectra taken in 1985--1990 and 2001 (Table~\ref{comphotphases}). During those epochs, a significant
difference of $\sim 3000$\,K in \teff~is noticed. In addition, a
remarkable difference is seen in \mdot~(which is 2 times higher in 2001 than in 1985--1990) and \vinf~(3 times lower
in 2001). We interpret the changes in \mdot~and \vinf~as a consequence of the
bistability of line-driven winds (Lamers et al. 1995). Indeed, there is a change in the iron ionization
structure around the sonic point, as predicted by Vink \& de~Koter (2002). The dominant iron ionization stage below the sonic point 
(which is around 18 \kms) changes from Fe$^{3+}$ in 1990 to Fe$^{2+}$ in 2001.

\begin{table}[!t]\centering
\setlength{\tabnotewidth}{\columnwidth}
\tablecols{5}
\caption{Stellar parameters during minimum } 
\label{comphotphases}
\begin{tabular}{l c c c c}
\toprule
Epoch & \rstar & \teff & \mdot & \vinf \\
&(\rsun) & (K) & (\msunyr) & (\kms)\\
\midrule
1985--1990  & 62  & 23,800 & $1.5 \times 10^{-5}$ & 300 \\
Apr 2001 & 69 & 20,500 & $2.5 \times 10^{-5}$ & 105 \\
\bottomrule
\end{tabular}
\end{table}

\subsection{The mass-loss rate evolution}

Figure~\ref{reff} (panel $a$) shows the mass-loss rate evolution along the S Dor cycle obtained from steady-state
and time-dependent models. As a result of the large flow timescale ($\sim 5$ years) during the cool phases,
significant differences are derived when time-dependent effects are taken into account. In special,
\mdot~is actually approximately constant during the cool phase (1991--1999), in comparison with a strong increase
in \mdot~around maximum when steady-state models are used. Thus, the results obtained from time-dependent
modeling imply
that the eruptions during the S Dor cycles begin well earlier than the maximum seen in the visual and near-infrared lightcurve. In addition, 
it is misleading the idea of an increase in \mdot~as a consequence of the maximum -- actually, the maximum in the
lightcurve is a consequence of the increase in \mdot~which occurred 2 or 3 years ago, in the beginning of the cool phase.

\subsection{Changes in the photospheric and hydrostatic radius}

\begin{figure}[!h]
  \includegraphics[width=\columnwidth]{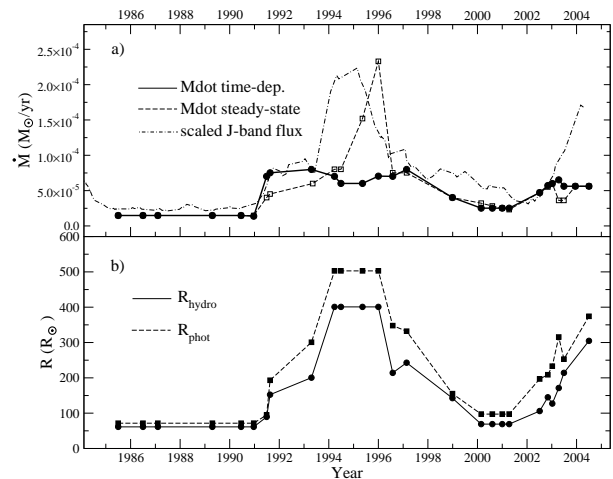}
  \caption{Evolution of the mass-loss rate (panel $a$) and of the hydrostatic and photospheric radius (panel $b$) 
  of AG~Car during the S Dor cycle.}
  \label{reff}
\end{figure}

The evolution of the photospheric and hydrostatic radius of AG Car during the S Dor cycle is presented in the
panel $b$ of Figure~\ref{reff}. It is apparent to conclude that {\em the S Dor-type variability is driven
mainly by changes in the hydrostatic radius of the star}. Changes in the photospheric radius do occur as a
result of the dense wind and variable mass-loss rate, but this change solely is not sufficient to explain the
strong temperature changes seen from minimum to maximum.

\end{document}